\documentclass[aps,twocolumn,superscriptaddress,showpacs]{revtex4}
\usepackage{graphicx}

\begin{document}

\title{All-optical active plasmonic devices with memory and power switching functionalities based on $\epsilon$-near-zero nonlinear metamaterials}

\author{Alessandro Ciattoni}
\affiliation{Consiglio Nazionale delle Ricerche, CNR-SPIN 67100 L'Aquila, Italy and Physics Department, University of L'Aquila, 67100 Coppito
(L'Aquila), Italy } \email{alessandro.ciattoni@aquila.infn.it}

\author{Carlo Rizza}
\affiliation{Electrical Engineering Department, University of L'Aquila, 67100 Zona industriale di Pile (L'Aquila), Italy}

\author{Elia Palange}
\affiliation{Electrical Engineering Department, University of L'Aquila, 67100 Zona industriale di Pile (L'Aquila), Italy}

\begin{abstract}
All-optical active plasmonic devices are of fundamental importance for designing efficient nanophotonic circuits. We theoretically propose and
numerically investigate an active plasmonic device made up of a nonlinear $\epsilon$-near-zero metamaterial slab of thickness smaller than $100$
nanometers lying on a linear $\epsilon$-near-zero metamaterial substrate. We predict that, in free-space coupling configuration, the device,
operating at low-intensity, would display plasmon mediated hysteresis behavior since the phase difference between the reflected and the incident
optical waves turns out to be multi-valued and dependent on the history of the excitation process. Such an hysteresis behavior would allow to regard
the proposed device as a compact memory unit whose state is accessible by measuring either the mentioned phase difference or the power, which is
multi-valued as well, carried by the nonlinear plasmon wave. Since multiple plasmon powers comprise both positive and negative values, the device
would also operate as a switch of the plasmon power direction at each jump along an hysteresis loop.
\end{abstract}

\maketitle

\section{INTRODUCTION}

Confinement and steering of optical radiation at nanometer scale are fundamental issues of modern nanophotonics and a dominant role is played by
surface plasmon polaritons \cite{Bronge,Shala1} (SPP) which are collective electron density oscillations coupled to external electromagnetic waves
\cite{Raethe} existing at metallo-dielectric interfaces. The surface charge oscillations merge the dielectric evanescent waves with the metal
exponentially decaying fields so that SSPs are surface modes tightly confined at the interface, a fundamental ingredient for the miniaturization of
optical components down to the sub-$100$-nm-size regime \cite{Barnes}. Aimed at designing photonic circuits, controlling SPPs flow is an important
topic and various waveguiding schemes have been proposed \cite{Ebbese,Gramot} together with a plenty of passive plasmonic devices encompassing
mirrors \cite{Ditlba}, lenses \cite{Liuuu1}, interferometers \cite{Ziaaaa}, splitters \cite{Bozhev} and resonators \cite{Minnnn}. Although passive
plasmonic devices can provide some of the essential components for signal processing, an efficient photonic circuitry amenable to be integrated with
nanoscale electronics evidently requires externally driven (active) plasmonic devices \cite{MacDon} and plasmonic emitters \cite{Koller,Hryciw},
detectors \cite{Tanggg,Neuten} and plasmon lasers \cite{Oulton,Kwonnn} have been considered. In order to manipulate the plasmon flow along the path
between the source and the detector, the dielectric refractive index near the metal surface has to be externally controlled and this can be done
through an electro-optic medium to obtain electrically-driven plasmonic modulators \cite{Caiiii}. Electro-optic methodology is particularly
intriguing since both optical and electrical signals can in principle be processed and interconnected on the same circuit board. Another viable way
to actively affect plasmon propagation consists in employing a nonlinear dielectric medium with a large Kerr nonlinearity and this has lead to the
observation of plasmon mediated optical bistability \cite{Wurtz1} and all-optical modulation of surface plasmon polaritonic crystal transmission
\cite{Wurtz2}. The main drawback to the use of the Kerr nonlinearity is that the nonlinear response is generally weak and, even if SSPs are
accompanied by a large field enhancement, the required optical intensities are very large. For this reason, even though there is a considerable
research effort in nonlinear plasmonics \cite{Palomb,Shaooo,Yinnnn,Marini}, an efficient nonlinear active plasmonic device has not been devised yet.

The research interest in surface plasmons has been further increased by the availability of metamaterials which are artificially structured materials
whose underlying subwavelength constituents are suitably tailored to allow the engineering of the overall medium electromagnetic response. As an
example, it has been recently suggested that the flow of SPPs on metallo-dielectric interfaces can be efficiently (and passively) controlled by
exploiting the metamaterial based transformation optics technique to devise the optical parameters of the dielectric medium \cite{Huidob,Liuuuu}.

The investigation of metamaterial nonlinear properties is particularly important in that it can lead to overcoming one of the fundamental limit of
nonlinear optics, the fact that most of the optical materials have a relatively weak nonlinear response. The main idea is that the local
electromagnetic fields of the inclusions in the metamaterial can be much larger than the average value of the field thus producing an enhancement of
the nonlinear response \cite{Pendr3}. Following a very different route, it has recently been proposed that an extremely marked nonlinear behavior can
be observed, instead of by enhancing the nonlinear response, by substantially reducing the metamaterial linear dielectric permittivity \cite{Ciatt1}
and this strategy has allowed to predict self power-flow reversing of an optical beam \cite{Ciatt2} and transmission optical bistability of a slab of
sub-wavelength thickness at very low optical intensities \cite{Ciatt3,Ciatt4}. It is worth noting that, as opposed to standard media where
nonlinearity requires macroscopic propagation lengths (greater than a centimeter) to produce an all-optical manipulation, $\epsilon$-near-zero
nonlinear metamaterials dramatically affect light just at their free-space interface (as discussed in Ref.\cite{Ciatt3}) and therefore they are
candidates to play a pivotal role in designing nanophotonic nonlinear active devices, one of which is discussed in the present paper.

In this paper we theoretically investigate a nanoscale all-optical active plasmonic device which, operating at low optical intensities and being
integrable in a photonic circuit, could be used both as plasmonic memory unit and as a device able to switch the direction of the power carried by a
surface plasmon wave. The system basically consists of a few tens of nanometers thick slab of a nonlinear Kerr metamaterial lying on a linear
metamaterial substrate, the nonlinear and linear medium having very small dielectric permittivities of opposite signs. Since the substrate
permittivity is much smaller than one, the critical angle to achieve phase-matching between free-space waves and surface plasmons is rather small (of
the order of few degrees) so that total reflection of a free-space plane wave impinging onto the nonlinear slab surface is easily observed and it is
accompanied by the excitation of a nonlinear plasmonic wave at the slab-substrate interface. The small value of the slab permittivity allows the
nonlinearity to produce a rather complex hysteresis behavior (in analogy of the transmissivity hysteresis described in Ref.\cite{Ciatt3}) which is
easily detectable by measuring the phase difference between the reflected and the incident optical wave. Therefore the system can be regarded as a
nanoscale memory unit on a board whose state is easily read out through a reflection experiment. Note that optical bistability from surface plasmon
excitation has been investigated employing metal-coated prisms in the attenuated total reflection (ATR) or Kretschmann geometry in the presence of a
nonlinear dielectric \cite{Wysinn,Hicker,Zhouuu} but these setups are not easily scaled down to be integrated in a photonic circuit due to the
macroscopic prism dimension, a difficulty completely avoided by the device discussed in this paper. The nonlinear surface plasmon carries optical
power along the slab-substrate interface and, due to the chosen extreme nonlinear regime, the overall plasmon power can be both parallel and
anti-antiparallel to the plasmon wave-vector, an exotic situation generally not occurring in the presence of linear materials with positive magnetic
permeability. The total plasmon power follows the hysteresis slab behavior and therefore the power direction can be switched simply by changing
either the input optical intensity or the incidence angle to force the system to jump to a different hysteresis state.

\section{Nonlinear Plasmonic Device}

We start by considering the plasmonic device consisting in a slab of thickness $L=84 \: nm$ lying on a substrate as reported in Figure 1a. Here we
consider monochromatic electromagnetic radiation whose free-space wavelength is $\lambda = 810 \: nm$. The slab is a nonlinear Kerr metamaterial with
a very small linear dielectric permittivity and a viable way to produce this artificial medium is to repeat along the $y$-axis nonlinear metal and
dielectric layers of thickness $d_1$ and $d_2$, respectively. If both thicknesses are much smaller than $\lambda$, radiation experiences a uniform
electromagnetic response obtained by averaging both the linear permittivities $\epsilon_1$ and $\epsilon_2$ and the nonlinear Kerr coefficients
$\chi_1^{(3)}$ and $\chi_2^{(3)}$ \cite{Ciatt1}.
\begin{figure}
\includegraphics[width=0.5\textwidth]{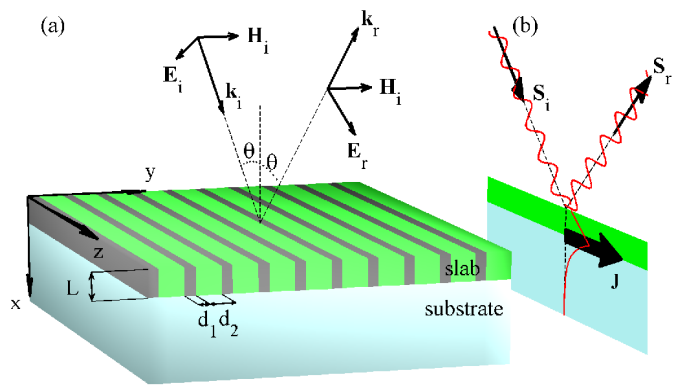}
\caption{(a) Geometry of the plasmonic device (slab and substrate) and of the free-space incident (i) and reflected (r) plane waves. (b) Sketch of
the electromagnetic device configuration. The incident and reflected waves are reported with red wavy lines while the arrows denoted with ${\bf S}_i$
and ${\bf S}_r$ represent their Poynting vectors. The nonlinear plasmon excited at the slab-substrate interface is reported with a red line and the
large arrow denoted with $J$ represents the overall power carried along the interface by the nonlinear surface mode.}
\end{figure}
Therefore, since metal layers have negative dielectric permittivity, it is possible to choose $d_1$ and $d_2$ in such a way that the slab average
dielectric permittivity is $\epsilon_{sl}=\langle \epsilon_i \rangle =-0.01$ ($\epsilon$-near-zero metaterial \cite{Silver,Aluuuu}) and $\chi^{(3)} =
\langle \chi_i^{(3)} \rangle >0$. Note that we are considering a real permittivity $\epsilon_{sl}$ and this can be achieved by using dielectric
layers with gain to balance the metal losses \cite{Ramak2}. The substrate is a linear metamaterial with dielectric permittivity $\epsilon_{su} =
0.01$ and it can be manufactured either by means of the same technique employed for the slab with linear layers or by exploiting the method discussed
in Ref.\cite{Zengg1,Zengg2}, consisting in dispersing silver coated spherical semiconductor quantum dots (with optical gain) in a homogeneous host
material whose permittivity can be externally tuned. As reported in Figure 1a, an incident plane wave (i) is made to impinge onto the slab-vacuum
interface (the plane $x=0$) with incidence angle $\theta$ and with the complex electric field ${\bf E}_i$ belonging to the plane of incidence
(parallel to the plane $xz$) which is parallel to the slab layers. A nonlinear plasmon mode localized at the slab-substrate interface (the plane
$x=L$) can be excited only if the incident plane wave is phase-matched with the substrate evanescent waves and this happens if $\theta$ is greater
than the critical angle $\theta_{cr} = \arcsin \sqrt{\epsilon_{su}} \simeq 5.7$ degrees. Therefore, for $\theta > \theta_{cr}$, total reflection
occurs since both $\epsilon_{sl}$ and $\epsilon_{su}$ are real and consequently $|{\bf E}_i|=|{\bf E}_r|= Q/ \sqrt{\chi^{(3)}}$ (where $Q$ is the
dimensionless electric field magnitude), the reflection solely yielding a phase difference $\Psi$ between the reflected and incident waves given by
\begin{figure}
\includegraphics[width=0.45\textwidth]{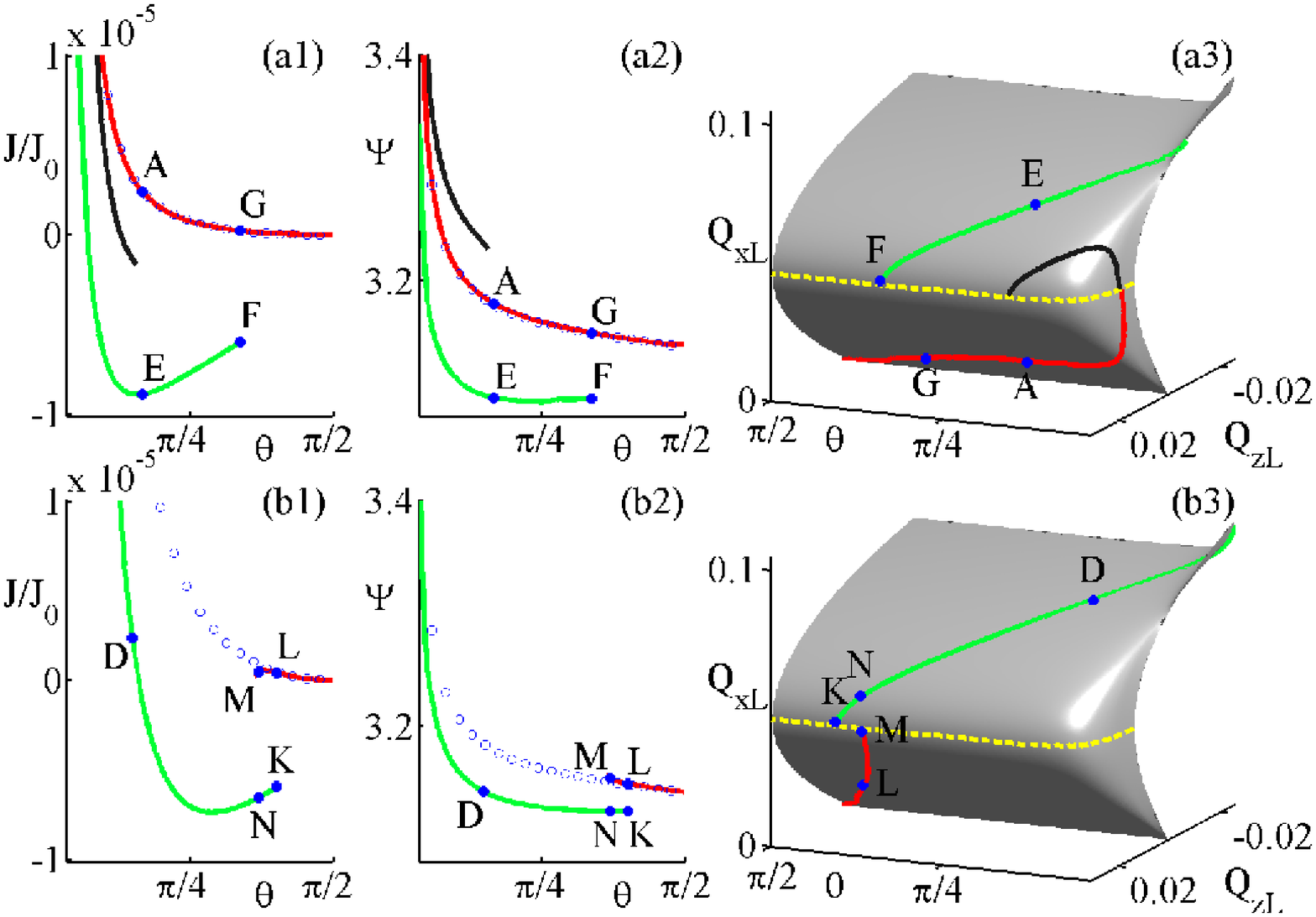}
\includegraphics[width=0.45\textwidth]{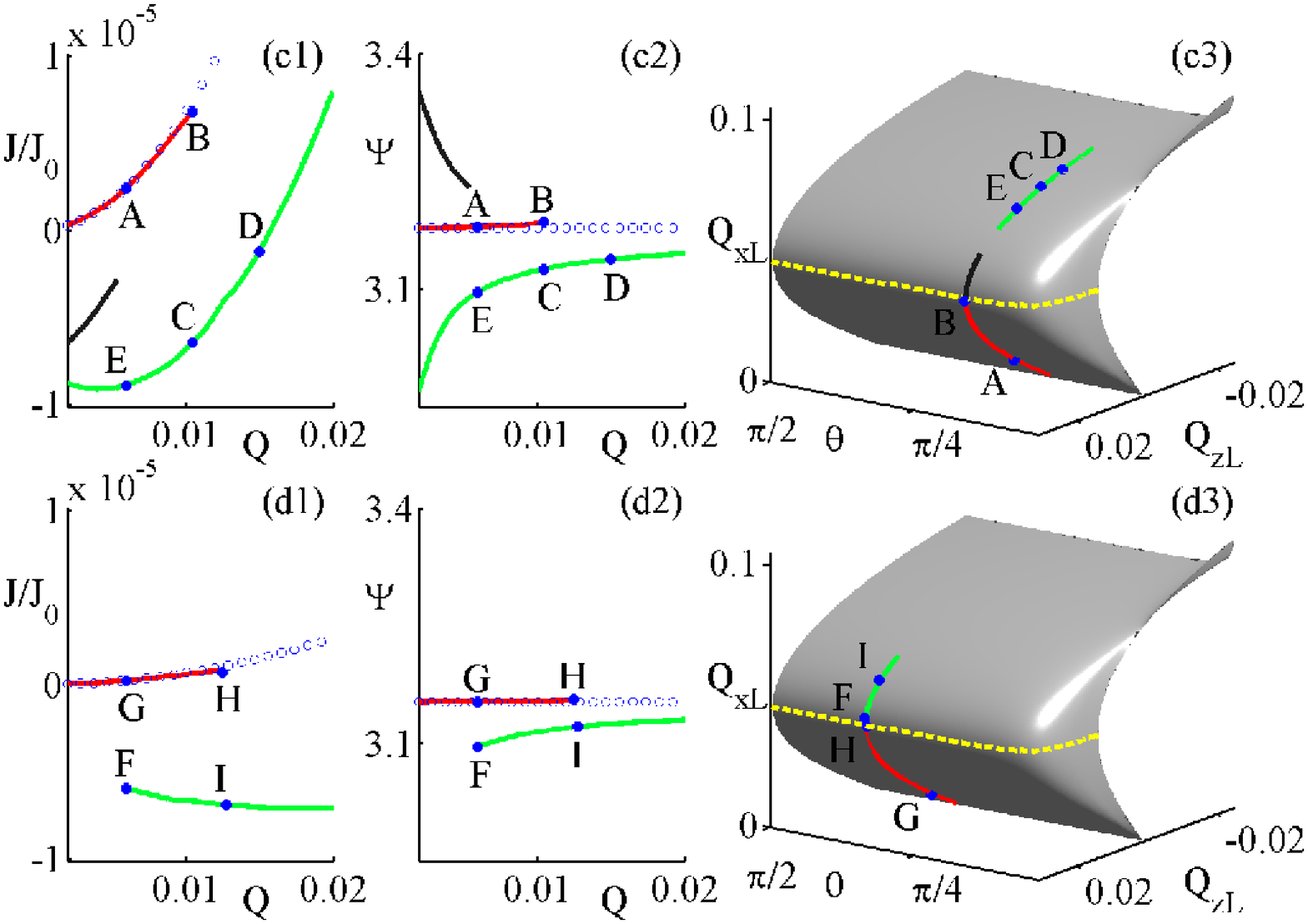}
\caption{Plots of the normalized power per unit length $J/J_0$ (here $J_0 = \sqrt{\epsilon_0 / \mu_0} /(2 k_0 \chi^{(3)})$) and phase difference
$\Psi$ along four possible ways of varying the excitation parameters. (a) $Q= 0.006$ (fixed) and $\theta_{cr} < \theta < \pi/2$ (variable); (b) $Q=
0.015$ (fixed) and $\theta_{cr} < \theta < \pi/2$ (variable); (c) $0 < Q < 0.02$ (variable) and $\theta = \pi/6$ (fixed); (d) $0 < Q < 0.02$
(variable) and $\theta = \pi/3$ (fixed). In each panel, the curves plotted by blue empty circles represent the linear counterparts of $J$ or $\Psi$.
The surface $Q_{xL}$ obtained by Eq.(\ref{surf}) is plotted in panels a3, b3, c3 and d3 together with the critical points curve reported with a
dashed yellow line.}
\end{figure}
\begin{equation} \label{Psi}
e^{i \Psi} = \frac{{\bf E}_r \cdot \hat{\bf e}_x}{{\bf E}_i \cdot \hat{\bf e}_x}.
\end{equation}
As schematically sketched in Figure 1b, total reflection produces a nonlinear localized mode at the interface $x=L$ which is a transverse magnetic
(TM) field (see Appendix A).  Evidently, the energy flow is purely along the $z$-direction since the Poynting vector is ${\bf S}= \frac{1}{2}
\textrm{Re} \left( {\bf E} \times {\bf H}^* \right) = S_z(x) \hat{\bf e}_z$ everywhere and hence, the total power per unit length carried by the
nonlinear wave is
\begin{equation} \label{J}
J = \int_0^\infty dx S_z (x).
\end{equation}

We have numerically evaluated nonlinear plasmon modes (see Appendix A) for a number of excitation states ($\theta$, $Q$) characterized by the
incidence angle $\theta$ and the incident plane wave amplitude $Q$ and, for each mode, we have evaluated the phase difference $\Psi$ from
Eq.(\ref{Psi}) and the power per unit length $J$ from Eq.(\ref{J}). In Figure 2 we report the evaluated $J$ and $\Psi$ along four possible ways
(denoted with a, b, c and d) of varying the parameters ($\theta$, $Q$). In situations (a) and (b) the amplitude is hold fixed with $Q = 0.006$ and $Q
= 0.015$, respectively and the incidence angle is varied in the range $\theta_{cr} < \theta < \pi/2$ whereas in situations (c) and (d) the angle is
hold fixed with $\theta = \pi/6$ and $\theta = \pi/3$, respectively and the amplitude is varied in the range $0< Q < 0.02$. For comparison purposes,
in each panel of Figure 2 reporting $J$ or $\Psi$ we have also plotted (blue empty circles curves) the corresponding {\it linear} quantities i.e. $J$
or $\Psi$ as evaluated for a linear slab of dielectric permittivity $\epsilon_{sl}$. Note that, in each situation reported in Figure 2, both $J$ and
$\Psi$ are multi-valued function of the excitation parameters or, more precisely, this quantities generally assume more than one value for a given
excitation state ($\theta$, $Q$). This implies that the considered system displays multistability controllable through the incidence angle and the
input optical intensity, and detectable by measuring $J$ or $\Psi$. From Figure 2, it is evident that the values of $J$ and $\Psi$ lie on at most
three different branches (red, green and black lines) and, most importantly, that the branches presents points at which they abruptly stop (see,
e.g., state $F$ of panels a1 and a2, states $M$ and $K$ of panels b1 and b2, state $B$ of panels c1 and c2 and states $F$ and $H$ of panels d1 and
d2). The fact that different branches abruptly stop and that they are not mutually continuously connected imply that the multistability discussed in
this paper is supported by a mechanism fundamentally different from that (feedback effect) yielding the standard optical bistability.

\section{Plasmonic Multistability}

In order to discuss the physical mechanism supporting the obtained multi-stable phenomenology it is convenient to consider the electromagnetic
matching conditions at the interface $x=L$ which yield
\begin{equation} \label{surf}
\left[ \epsilon_{sl} + \frac{1}{2} \left( 3 Q_{xL}^2 + Q_{zL}^2 \right) \right] Q_{xL} + \epsilon_{su} \sqrt{\frac{\sin^2 \theta}{\sin^2 \theta -
\epsilon_{su}}} Q_{zL} =0
\end{equation}
where $Q_{xL}=Q_x (L)$ and $Q_{zL}=Q_z (L)$ are the values of the $x$ and $z$ dimensionless electric field components at the interface $x=L$ on the
slab side (see Appendix A). For $\theta$ and $Q_{zL}$ assigned, Eq.(\ref{surf}) provides the allowed values of $Q_{xL}$ (electric field component
normal to the slab-substrate interface). Note that for a linear slab the first term of Eq.(\ref{surf}) reduces to $\epsilon_{sl} Q_{xL}$ this
implying that only one value of $Q_{xL}$ is allowed. In the four panels a3, b3, c3 and d3 of Figure 2 we plot the surface $Q_{xL}$ obtained from
Eq.(\ref{surf}) in the range $Q_{xL}>0$ (without loss of generality) and we note that such surface is folded. This implies that, due to the effect of
the nonlinearity, the electromagnetic boundary conditions generally allow more than one electromagnetic configuration at a given $\theta$ and plasmon
amplitude $Q_{zL}$, the various configurations having different values of $Q_{xL}$. Each point of the surface corresponds to a nonlinear plasmon mode
and therefore, in panels a3, b3, c3 and d3 of Figure 2, we have also reported the various state curves of plasmon modes used to evaluate $J$ and
$\Psi$ of the corresponding panels in the Figure, using the colors to establish the correspondence. The red branches of both $J$ and $\Psi$ of Figure
2 practically coincide with their linear counterparts (the empty circles curves) apart from the region close to the breaking points so that we
conclude the green branches to be of pure nonlinear origin and arising from the surface folding which is therefore the crucial element producing the
multistability of $J$ and $\Psi$. It is worth noting that folding takes place if the surface admits critical points where the partial derivative of
Eq.(\ref{surf}) with respect $Q_{xL}$ vanishes, i.e.
\begin{equation} \label{crit}
\epsilon_{sl} + \frac{1}{2} \left( 9 Q_{xL}^2 + Q_{zL}^2 \right) =0,
\end{equation}
which represents the curve of critical points reported with a dashed yellow line in panels a3, b3, c3 and d3 of Figure 2. From Eq.(\ref{crit}) we
readily deduce that critical points can appear only if the field amplitudes $|Q_{xL}|$ and $|Q_{zL}|$ are comparable with $\sqrt{-\epsilon_{sl}}$ and
this physically means that linear and nonlinear contributions to the overall electromagnetic slab response must be comparable. This is an extreme
nonlinear condition \cite{Ciatt1} made here possible by the fact that the slab permittivity $\epsilon_{sl}$ is much smaller than one. It is worth
noting that, in a conventional nonlinear medium where the permittivity is of the order of unity, the above folding would required an optical
intensity so large to effectively prevent the whole discussed phenomenology to take place. We therefore conclude that the multistability discussed in
this paper can uniquely be observed by means of $\epsilon$-near-zero nonlinear metamaterials. To get further physical insight, note that the
constitutive relations yield $D_x (L,0) = \epsilon_0 \left[ \epsilon_{sl} + \frac{1}{2} \left( 3 Q_{xL}^2 + Q_{zL}^2 \right) \right]
Q_{xL}/\sqrt{\chi^{(3)}}$, so that Eq.(\ref{crit}) coincides with the condition $ \partial D_x(L,0) / \partial Q_{xL} =0$. Therefore the plasmon
states below and above the critical curve are characterized by $\partial D_x(L,0) / \partial Q_{xL}  >0$ and $ \partial D_x(L,0) / \partial Q_{xL}
<0$, respectively and therefore characterized by very different electromagnetic features (the curve of critical point playing the role of a
separatrix between the two electromagnetic regimes). This is particularly evident from panel d3 of Figure 2 where states $H$ and $F$ are just below
and above the critical curve, respectively (and hence their electric amplitudes are almost equal at $x=L$) and nonetheless their corresponding values
of $J$ and $\Psi$ of panels d1 and d2, respectively, are very different. The existence of two distinct electromagnetic regimes explains both the
different values of $J$ and $\Psi$ along the red and green branches (bistability) and branch breaking since, from Figure 2, it is evident that the
points at which the branches of $J$ and $\Psi$ abruptly stop correspond to a plasmon mode whose $Q_{xL}$ and $Q_{zL}$ lie just either below or above
the yellow critical points curve (state $F$ of panels a1, a2 and a3, state $M$ and $K$ of panels b1, b2 and b3, etc.)

\section{Device memory and power switching functionalities}

By combining the discussed multistability with the branch breaking it is easy to prove that the considered system displays a complex hysteresis
behavior allowing to regard the system as a device with structured memory functionality. Suppose, for example, that the incident plane wave is such
that $\theta=\pi/3$ and $Q=0.001$ so that, from panels d1, d2 and d3 of Figure 2, it is evident that a state of the red curve is excited since for $Q
< 0.004$ the green branch has no corresponding allowed states. By increasing the input intensity (and hence $Q$) the state $G$ is reached and
surpassed toward the state $H$. Once $H$ is reached, if the input intensity is further increased the system has to undergo a sudden jump to the state
$I$ belonging to the green branch since the red branch stops at $H$. From the state $I$ the input intensity can now be decreased toward the state $F$
and therefore along such a backward path the attained values of $J$ and $\Psi$ are different from the forward path, i.e. hysteresis occurs.
Evidently, by decreasing the input intensity from the state $F$ the system undergoes another jump to reach the state $G$, thus coming back to the red
branch and closing the hysteresis loop. The system hysteresis phenomenology is however far more rich since the incidence angle can also be varied.
Suppose, for example, that the incident plane wave is such that $\theta=\pi/6$ and $Q=0.015$ so that, from panels b1, b2 and b3 of Figure 2, it is
evident that the state $D$ belonging to the green branch is excited since the red branch has no corresponding allowed state. If the angle is
increased, by reasoning in analogy with the just discussed hysteresis loop, the system can be brought to the state $K$ (passing through the state
$N$) and hence forced to jump to the state $L$ since $K$ is a breaking point of the green branch.
\begin{figure}
\includegraphics[width=0.45\textwidth]{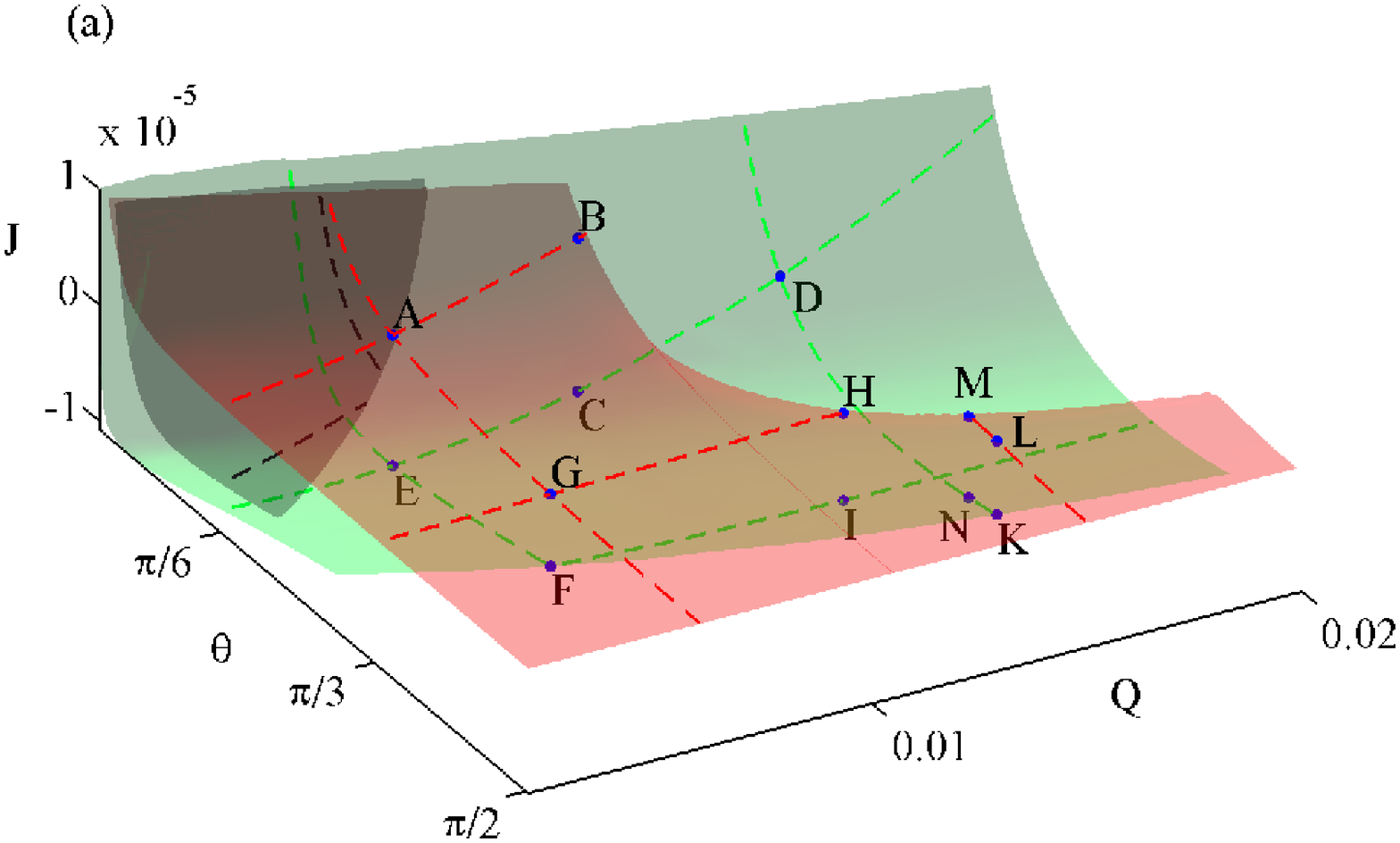}
\includegraphics[width=0.45\textwidth]{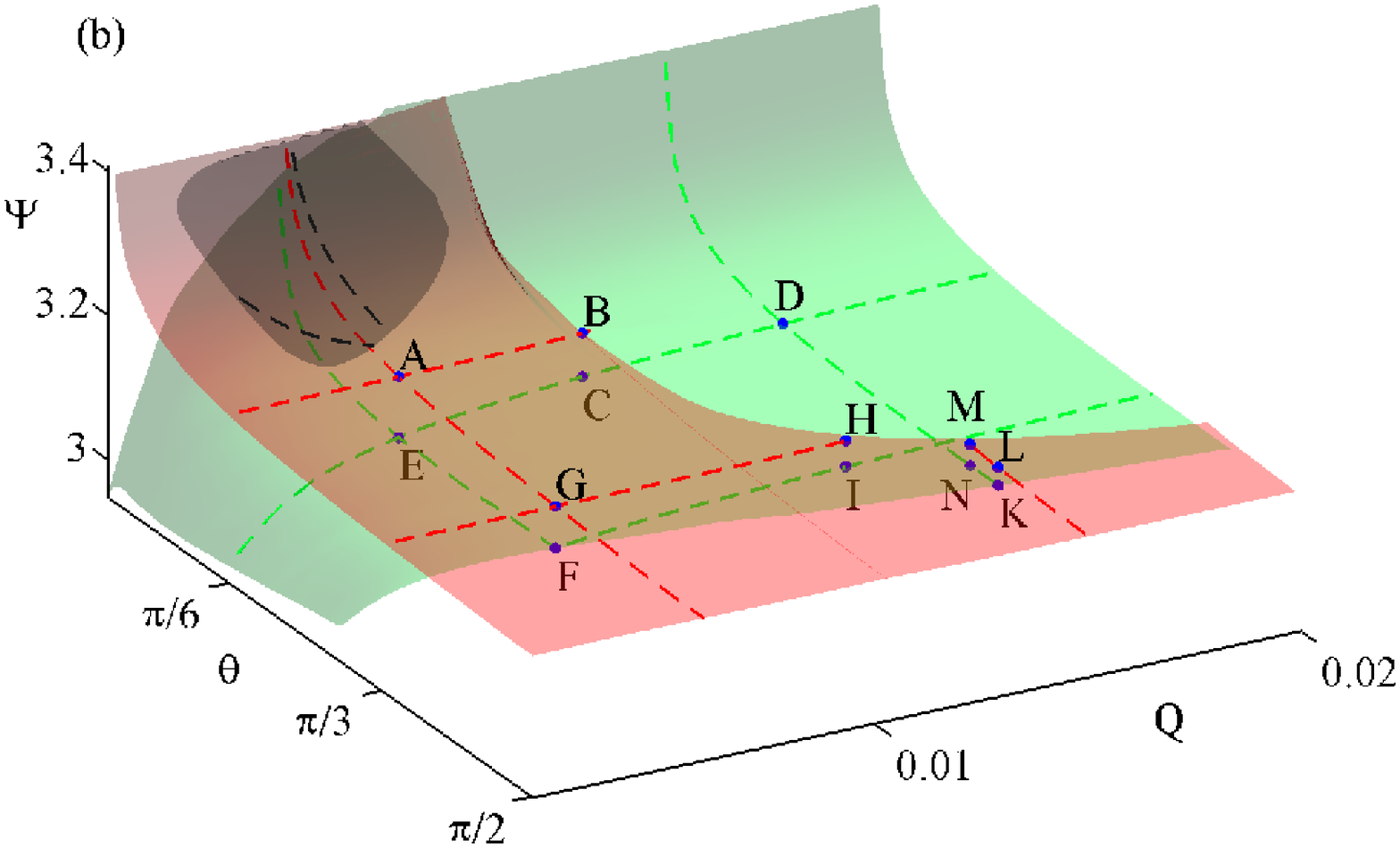}
\caption{(a) Global multi-valued surface of the normalized power per unit length $J/J_0$ (here $J_0 = \sqrt{\epsilon_0 / \mu_0} /(2 k_0
\chi^{(3)})$). (b) Global multi-valued surface of the phase difference $\Psi$. Dashed lines represent the branches of $J$ and $\Psi$ considered in
Figure 2 and analogously for the states reported with capital letters.}
\end{figure}
From $L$ the system can be driven to $M$ by decreasing the incidence angle and eventually it can be forced to jump down to the state $N$ thus closing
the "angular" hysteresis loop.

The number of possible hysteresis loops is infinite since $\theta$ and $Q$ can independently be varied and, in order to provide a global description
of the system multistable behavior we report in panels (a) and (b) of Figure 3 the values of $J$ and $\Psi$, respectively, for all the possible
$\theta$ and $Q$ in the ranges $\theta_{cr} < \theta < \pi/2$ and $0 < Q < 0.02$. In Figure 3 we also plot, through dashed lines and corresponding
colors, all the branches of $J$ and $\Psi$ considered in Figure 2 together with all the states denoted with capital letters. The surfaces
representing $J$ and $\Psi$ exhibit more than one sheet, each one admitting breaking lines, and the mutual position of the just discussed hysteresis
loops G-H-I-F and N-K-L-M is particularly evident. If we choose to operate uniquely with the red and green sheets, it is evident that the system is
suitable to record a binary information which can be easily read since a single measure of $\Psi$ (or $J$) allows to know whether a red or a green
state is excited. Moreover, the state of the memory unit can be simply altered by driving the system through a breaking line thus forcing a switch of
its operating sheet. Note also that the system memory functionality can even be suitably tailored to satisfy external requirements (i.e. specific
optical intensities or angles imposed by a possible nearby circuit environment) since, for each plasmonic state, a hysteresis loop can be found which
starts and ends at the considered state. Suppose, as an example, that an hysteresis loop containing the red state $A$ of Figures 2 and 3 is required.
Once the state A is excited, one can increase the field amplitude $Q$ to reach the state B (which is a red sheet breaking point) and to force the
system to jump to the green state $C$. From panel c1 and c2 of Figure 2 or from Figure 3 it is apparent that by solely varying the field amplitude
the system can never be brought back to the state A. However, one can drive the system to reach the state E, fix the field amplitude Q and increase
the angle $\theta$ to reach the state F (see panel a1 and a2 of Figure 2 or Figure 3) which is a breaking point of the green sheet. Therefore a jump
can be induced to make the system reach the red state G and, after reducing the angle, to come back to the initial state A thus closing the
hysteresis loop.

In addition to the above discussed multistability and related memory functionality, we note that, from Figures 2 and 3, both positive and negative
signs of $J$ are allowed. Specifically, the red branches of $J$ are always positive whereas the green branches can be both positive and negative.
Bearing in mind that the red branches (where they exist) almost coincides with the slab linear behavior, we conclude that the extreme nonlinear
regime also introduces the possibility of an overall plasmon power which is antiparallel to the wave-vector. In fact, since the slab and substrate
permittivities $\epsilon_{sl}$ and $\epsilon_{su}$ are negative and positive, respectively, the Poynting vector is locally antiparallel and parallel
to the $z$-direction (which coincide with the plasmon mode wave-vector) and hence the total power can be both positive and negative. For the red
branches of $J$ of Figure 2 (or 3) the positive contribution to $J$ coming from the evanescent field in the substrate overcomes the slab negative
contribution thus producing an overall positive $J$. However, the plasmon modes belonging to the green branches are generally characterized by a
larger amount of electromagnetic energy stored within the slab since their amplitudes $Q_{x}$ are generally larger than their red counterparts as it
is evident from panels a3, b3, c3 and d3 of Figure 2. Therefore the negative contribution to $J$ of the green modes can overcome the positive
evanescent contribution thus yield an overall negative plasmon power. As an example, states A and E of panels a1 of Figure 2 are excited through the
same input intensity and incidence angle and, nonetheless, $J(A)>0$ and $J(E)<0$. Obviously, for small values of $\theta$, the extension of the
evanescent field thorough the substrate is so large to always yield a positive value of $J$ (see the positive portion of the green sheet of $J$ in
Figure 3a).

The fact that the overall plasmon power, for a given excitation state $(\theta,Q)$, can be both parallel or anti-parallel to the wave vector also
allows to regard the considered system as a device able to switch the direction of the electromagnetic energy flowing along the slab-substrate
interface. In fact, once a state is excited with a given sign of $J$, the energy flow direction switch is easily performed by externally changing the
incident plane wave and driving the system through a sheet breaking line whose crossing forces the system to jump to the other sheet where the sign
of $J$ is generally different from the initial one.

\section{Device feasibility and design}

The feasibility of the above described active plasmonic device is evidently related to the actual availability of both slab and substrate
metamaterials exhibiting the above mentioned features. In order to prove that such media can be actually synthesized, we have chosen $d_1 = 2$ nm,
$d_2 = 5$ nm, $L = 84$ nm, $\epsilon_1 = -28.81 + 9.78 i$, $\epsilon_2 = 11.51 - 3.91 i$, $\chi_1^{(3)} = 3.16 \times 10^{-16} \: m^2/V^2$ and
$\chi_2^{(3)} = 0$. The parameters of medium $1$ coincide with those of silver \cite{Palikk}, characterized by a very large nonlinear susceptibility
\cite{YangGu}, with the imaginary part of the permittivity corrected by the layer size effect (since $d_1 = 2$ nm) \cite{CaiSha}. Medium $2$ is a
linear dielectric with gain (to compensate the metal losses) and its permittivity $\epsilon_2$ can be achieved, for example, by embedding, in a
polymer background, a bi-dimensional array of colloidal II-VI quantum dots which are properly electrically or optically pumped \cite{Fuuuuu} through
ultra-violet light. The value of $\epsilon_2$ here considered is obtained through the Maxwell Garnett mixing rule where $\epsilon_b=10.2$ and
$\epsilon_{QD} = 10 - 15 i$ \cite{Fuuuuu} are the background polymer and quantum dots permittivities, respectively and $f=0.29$ is the quantum dots
volume filling fraction; besides both the polymer and the quantum dots have nonlinear susceptibility much smaller than $\chi_1$ so that the
requirement $|\chi_2| \ll |\chi_1|$ is assured. Note that, even though the actual fabrication of the considered ultra-small structure is hindered by
the very small thickness/width (i.e. $d_1/L$) ratio of the metallic grooves, the required array of metallic nanowires can be manufactured through the
general method of Superlattice Nanowire Pattern Transfer (SNAP) \cite{MeloshB,JungJoh,Martinez} which is a novel deposition technique particularly
suitable for producing ultrahigh-density arrays of aligned metal and semiconductor nanowires and nanowire circuits. Once the array of silver
nanowires is deposited onto the substrate, the vacancies between the metallic grooves can be filled with the required quantum dots dispersed in a
suitable liquid monomer (by using, for example, the standard spin-deposition process) which can be eventually polymerized through UV curing. The
effective (average) permittivity and nonlinear susceptibility of the considered sample are $\epsilon_{sl} = (d_1 \epsilon_1 + d_2 \epsilon_2)/(d_1 +
d_2)= -0.01 + 0.001i$ and $\chi^{(3)} = (d_1 \chi_1^{(3)} + d_2 \chi_2^{(3)})/(d_1 + d_2) = 9 \times 10^{-17} \: m^2/V^2$. As far as the substrate,
we choose here to exploit the just proposed scheme of dispersing silver (for which $\epsilon_{Ag} = -28.81 + 9.78 i$) coated spherical semiconductor
quantum dots (the same colloidal II-VI quantum dots as in the slab for which $\epsilon_{QD} = 10 - 15 i$) in a homogeneous host material (for which
$\epsilon_h = 10.2$). Exploiting the electrostatic treatment of Ref. \cite{Zengg1} and choosing $\rho = 0.19630$ for the fraction of the total
particle volume occupied by the inner quantum dot core we obtain for the effective permittivity of the overall inclusion $\epsilon_{in} = -17.87 +
0.05i$. Exploiting the standard Maxwell Garnett mixing rule and choosing $f= 0.04625$ for the volume filling fraction of the inclusions dispersed in
the hosting matrix, we obtain for the effective dielectric permittivity of the substrate $\epsilon_{su} = 0.01244 + 0.00001 i$.

\section{Full-wave simulations}

In order to test the behavior of the above discussed active plasmon device we have performed 3D full-wave simulations by means of a finite-element
solver (see Appendix B) in which we have used the numerical values of the parameters just chosen in the above structure design. In panel (a) of
Figure 4 we report the comparison between the phase difference $\Psi$ evaluated through full wave simulations (blue circles) for $\theta = \pi /6$
and $0 < Q < 0.025$ and the corresponding phase difference $\Psi$ of panel (c2) of Figure 2 (red and green solid lines).
\begin{figure}
\includegraphics[width=0.45\textwidth]{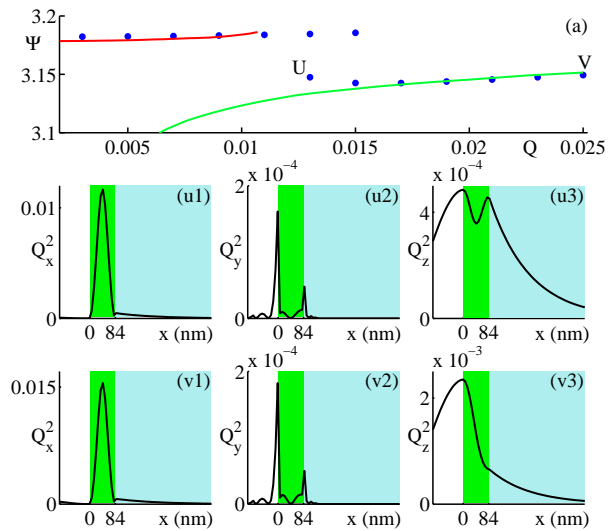}
\caption{(a): Comparison between the phase difference $\Psi$ evaluated through full wave simulations (blue circles) for $\theta = \pi /6$ and $0 < Q
< 0.025$ and the corresponding phase difference $\Psi$ of panel (c2) of Figure 2 (red and green solid lines). (u1), (u2), (u3) and (v1), (v2), (v3):
profiles of the plasmonic squared dimensionless electric field components as a function of $x$ (slab: $0<x<84$ nm; substrate: $x>84$ nm)
corresponding to the states $U$ and $V$ reported in panel (a). The electric field profiles have been evaluated at $y \simeq d_1$, e.g. close to the
metallic-dielectric interface.}
\end{figure}
We note that full-wave simulations confirm the existence of the above discussed multistability and branch breaking. For values of $Q$ far away from
the breaking value $Q=0.01$, full-wave simulations reproduces the predictions of the homogenized approach. On the other hand, discrepancies appear
for $0.01 < Q < 0.015$ and, specifically, the breaking point of the upper branch predicted by the full-wave simulations is at $Q=0.015$ instead of
$Q=0.01$ of the homogenized theory. The origin of the discrepancies and of the breaking point shift is due to the existence of a further plasmonic
resonance arising from the fact that a transverse magnetic plane wave (polarized in the $xz$ plane, see Figure 1a) is made to impinge onto a
metallo-dielectric medium layered along the $y$-direction. Therefore an additional $y$ component of the electromagnetic field of plasmonic origin
arises thus departing the layered medium behavior from that of its homogeneous counterpart. In order to support this reasoning we plot in panels
(u1), (u2) and (u3) and panels (v1), (v2) and (v3) of Figure 4, the profiles of the squared dimensionless electric field components as a function of
$x$ corresponding to the states $U$ and $V$ reported in panel (a), respectively. Such squared electric field profiles have been evaluated near the
metallic-dielectric interface ($y \simeq d_1$) where the additional plasmonic states are mainly localized. The state $V$ does not display significant
discrepancies between full-wave simulations and homogenized approach prediction (see panel (a) of Figure 4) since, as it is evident from panels (v1),
(v2) and (v3) of Figure 4, both the $x$ and $z$ electric field components are much larger than the $y$ component. On the other hand, the discrepancy
accompanying the state $U$ are due to the fact that the $y$ electric field component of panel (u2) is comparable with the $z$ component of panel
(u3).

The intensity of the incident plane wave is $I=(1/2)\sqrt{\epsilon_0 / \mu_0} Q^2/\chi^{(3)}$ which, for the amplitude range $Q < 0.02$ (where
multistability occurs for $\epsilon_{sl} = -0.01$) and for the above effective nonlinear susceptibility $\chi^{(3)}$, yields $I < 1.17 \: MW / cm^2$,
which are intensities smaller than those normally required for observing the standard optical bistability. However it is evident that the smaller
$|\epsilon_{sl}|$ the smaller the intensity required for observing the device multi-stable behavior so that very smaller intensity thresholds (of the
order of $W/cm^2$) are very likely to be attained.

\section{CONCLUSIONS}

In conclusion we have theoretically considered an all-optical active plasmonic device characterized by a complex multi-stable behavior allowing the
device, in principle, to be used both as a memory unit or as a switch of the plasmonic power direction. Since the transverse dimension of the
proposed device active region (slab) is smaller than $100$ nm and the system is controlled through free-space coupling, the device could be easily
integrable on a photonic circuit where it can be possibly interconnected with other nanophotonic and/or microelectronic devices. As opposed to
standard situations where the nonlinearity mediated plasmonic flow steering is hampered by the weakness of the nonlinear response and the short
propagation distances, the nanophotonic device proposed in this paper fruitfully exploits the nonlinear response since the linear dielectric
permittivities are very small and, consequently, nonlinearity rules the electromagnetic behavior just at the slab-substrate interface. We believe
that our methodology offers a new way for exploiting optical nonlinearity to actively control plasmonic propagation and hence for the designing of
future active plasmonic devices with complex functionalities.

\appendix
\section{Nonlinear Plasmonic Modes}

We have numerically investigated the excitation of nonlinear plasmonic modes at the slab-substrate interface by solving the slab nonlinear Maxwell
equations together with the electromagnetic matching conditions at $x=0$ and $x=L$. Specifically, the TM electromagnetic field within the slab is
${\bf E}= \left[Q_x(x) \hat{\bf e}_x + i Q_z(x) \hat{\bf e}_z\right] e^{i (k_0 \sin \theta) z} / \sqrt{\chi^{(3)}}$, ${\bf H} = \sqrt{\epsilon_0 /
\mu_0}\left[ Q_y(x) \hat{\bf e}_y \right]e^{i (k_0 \sin \theta) z} / \sqrt{\chi^{(3)}}$ (where $k_0 = 2 \pi / \lambda$ and $Q_x$, $Q_y$ and $Q_z$
are, without loss of generality, dimensionless amplitudes) and it satisfies Maxwell equations with the Kerr nonlinear constitutive relation ${\bf D}
= \epsilon_0 \left\{ \epsilon_{sl} {\bf E} + \chi^{(3)} \left[\left({\bf E} \cdot {\bf E}^* \right) {\bf E}  + \frac{1}{2} \left({\bf E} \cdot {\bf
E} \right) {\bf E}^* \right] \right\}$ for $0<x<L$, or
\begin{eqnarray}  \label{Maxwell}
\frac{1}{k_0} \frac{d Q_z}{dx} &=& Q_x \sin \theta - Q_y, \nonumber \\
\frac{1}{k_0} \frac{d Q_y}{dx} &=& \left[ \epsilon_{sl}+ \frac{1}{2} \left(Q_x^2 + 3 Q_z^2\right) \right] Q_z , \nonumber \\
Q_y \sin \theta &=& \left[ \epsilon_{sl} + \frac{1}{2} \left( 3 Q_x^2 + Q_z^2 \right)  \right] Q_x.
\end{eqnarray}
The free-space ($x<0$) incident and reflected plane waves are ${\bf E}_s = E_s (\hat{\bf e}_y \times {\bf k}_s/k_0) \exp \left(i {\bf k}_s \cdot {\bf
r}\right)$, ${\bf H}_s = E_s \sqrt{\epsilon_0/\mu_0} \hat{\bf e}_y \exp \left(i {\bf k}_s \cdot {\bf r}\right)$ (where $s=i,r$, $E_i$, $E_r$ are the
amplitudes of the two waves and ${\bf k}_i= k_0 \left(\cos \theta \hat{\bf e}_x + \sin \theta \hat{\bf e}_z \right)$, ${\bf k}_r= k_0 \left( - \cos
\theta \hat{\bf e}_x + \sin \theta \hat{\bf e}_z \right)$ are the wave vectors) whereas the evanescent field within the substrate ($x>L$) is
\begin{eqnarray}
{\bf E} &=& E_p \left(\sin \theta \hat{\bf e}_x - i \sqrt{\sin^2 \theta - \epsilon_{su}} \hat{\bf e}_z \right) \nonumber \\
        &\times& e^{- k_0 \sqrt{\sin^2 \theta - \epsilon_{su}} (x-L)} e^{i (k_0 \sin \theta) z}, \nonumber \\
{\bf H} &=& E_p \left( \epsilon_{su} \sqrt{\frac{\epsilon_0}{\mu_0}} \hat{\bf e}_y \right) e^{- k_0 \sqrt{\sin^2 \theta - \epsilon_{su}} (x-L)} e^{i
(k_0 \sin \theta) z} \nonumber \\ \label{evane}
\end{eqnarray}
where $E_p$ is, without loss of generality, the real field amplitude and $\theta > \theta_{cr}$ is assumed. At the slab input ($x=0$) and output
($x=L$) interfaces we require the continuity of the tangential electric field component ($E_z$) and of the normal displacement field component
($D_x$) (the continuity of the tangential magnetic field component $H_y$ evidently follows from the third of Eqs.(\ref{Maxwell})). In order to
evaluate a nonlinear plasmonic mode we specify the amplitude $E_p$ so that the matching conditions at $x=L$ yield
\begin{eqnarray} \label{bound}
Q_{zL} &=& - \sqrt{\sin^2 \theta - \epsilon_{su}} \sqrt{\chi^{(3)}}  E_p  \nonumber \\
\left[ \epsilon_{sl} + \frac{1}{2} \left( 3 Q_{xL}^2 + Q_{zL}^2 \right)  \right] Q_{xL} &=& \epsilon_{su} \sin \theta \sqrt{\chi^{(3)}}  E_p
\end{eqnarray}
where $Q_{xL}=Q_x (L)$ and $Q_{zL}=Q_z (L)$ are the values of the $x$ and $z$ dimensionless electric field components at the interface $x=L$ on the
slab side. After $Q_{xL}$ and $Q_{zL}$ are obtained from Eqs.(\ref{bound}), Eqs.(\ref{Maxwell}) are numerically solved all the way to the input face
at $z=0$ where the matching conditions allow to evaluate the amplitudes of the incident and reflected waves, $E_i$ and $E_r$, which turn out to have
the same absolute value. Note that, by combining the first and the second of Eqs.(\ref{bound}), Eq.(\ref{surf}) is easily obtained.

\section{Full-wave simulations}

The phase difference $\Psi$, plotted as blue circles in Figure 4(a), together with the profiles of the plasmonic squared dimensionless electric
components, reported in Figure 4, are obtained from a 3D full-wave simulations using the COMSOL RF Module. In order to perform the simulations
corresponding to the setup of Figure 1, we have considered an integration domain containing a unit cell consisting of a vacuum layer (in the region
$-160 <x<0 \: nm$, $0<y<7 \: nm$, $0<z<1 \: nm$), a metal layer characterized by $\epsilon_1=-28.81+9.78i$, $\chi_1^{(3)}=3.16\cdot 10^{-16} m^2/V^2$
(in the region $0 <x<84 \: nm$, $0<y<d_1=2 \: nm$, $0<z<1 \: nm$), a gain medium with $\epsilon_2=11.51-3.91i$, $\chi_2^{(3)}=0$ (in the region $0
<x<84 \: nm$, $d_1<y<d=7 \: nm$, $0<z<1 \: nm$) and a homogeneous substrate with $\epsilon_{su}=0.01$ ($84 <x<1704 \: nm$, $0<y<7 \: nm$, $0<z<1 \:
nm$). Periodic boundary conditions have been imposed at the faces perpendicular to the $y-$axis (at $y=0$ and $y=7 \: nm$) for achieving an infinite
perfect tiling of the system whereas periodic Floquet boundary conditions have been imposed at the faces perpendicular to the $z-$axis (at $z=0$ and
$z=1 \: nm$) for adapting the simulation domain to the $z$ field periodicity. Scattering boundary conditions have been imposed at the faces
perpendicular to the $x-$axis (at $x=-160 \: nm$ and $x=1704 \: nm$) for both exciting the incident plane wave and to absorb the waves scattered by
the structure. The simulation was performed by using a rectangular nonuniform mesh adapted to the structure design and the considered radiation
excitation.

\end{document}